\documentstyle[prl,eqsecnum,graphics,aps]{revtex}
\newcommand{\equn}[1]{\begin{equation}\label{#1}}
\newcommand{\eqan}[1]{\begin{eqnarray}\label{#1}}
\newcommand{\eqa}{\begin{eqnarray}}
\newcommand{\equ}{\begin{equation}}
\newcommand{\nuqe}{\end{equation}}
\newcommand{\uqe}{\end{equation}}
\newcommand{\naqe}{\end{eqnarray}}
\newcommand{\aqe}{\end{eqnarray}}
\newcommand{\nonu}{\nonumber}

\newcommand{\scs}{\scriptstyle}

\newcommand{\goto}{\rightarrow}
\newcommand{\half}{\frac{1}{2}}

\newcommand{\n}{\nabla_\perp}
\newcommand{\e}{{\rm e}}
\newcommand{\ab}{\alpha \beta}
\begin{document}

\twocolumn[\hsize\textwidth\columnwidth\hsize\csname@twocolumnfalse%
\endcsname
\title{Corrugation-Induced First-Order Wetting:\\
An Effective Hamiltonian Study}
\author{P.\ S.\ Swain \cite{email}}
\address{Max-Planck Institut f\"{u}r Kolloid- und Grenzfl\"{a}chenforschung,\\
Kantstrasse 55, D-14513 Teltow-Seehof, Germany}
\author{A.\ O.\ Parry}
\address{Department of Mathematics, Imperial College,\\ 180 Queen's Gate, London SW7 2BZ, United Kingdom.}
\draft
\maketitle

\begin{abstract}   
We consider an effective Hamiltonian description of critical wetting transitions in systems with short-range forces at a corrugated (periodic) wall. We are able to recover the results obtained previously from a `microscopic' density-functional approach in which the system wets in a {\it discontinuous} manner when the amplitude of the corrugations reaches a critical size $A^*$. Using the functional renormalization group, we find that $A^*$ becomes dependent on the wetting parameter $\omega$ in such a way as to decrease the extent of the first-order regime. Nevertheless, we still expect wetting in the Ising model to proceed in a discontinuous manner for small deviations of the wall from the plane.
\end{abstract}

\vskip2pc]

\section{Introduction}
In a recent article \cite{us} we have studied wetting transitions at a periodic non-planar substrate (i.e.\ a corrugated wall) within a Landau-like mean-field (MF) square gradient theory, which is appropriate for modelling adsorption in fluid and simple magnetic systems with short-range forces \cite{dietrich}. The surprising conclusion of this analysis was that second-order (critical) wetting transitions occurring in the planar system are generically corrugation-induced first-order when the root mean square amplitude $A$ of the corrugations exceeds a tricritical value $A^*$, whose numerical value is less than a bulk correlation length. The analysis of the Landau model is based on a perturbative expansion of the free energy about the equilibrium value of the planar system. It transpires that the free energy correction  due to corrugation can be expressed in terms of (planar) correlation functions, and is amenable to a graphical interpretation which generalizes the well known Cahn construction \cite{telo} for the planar system. Despite the convenience of a geometrical description and the simplicity of the final results, the calculation is rather  involved and perhaps obscures the underlying mechanism for the shift in the order of the transition. In this paper, we seek to elucidate the simplest possible effective Hamiltonian theory consistent with the previously derived results for the shifted MF phase boundary and proceed to discuss the influence of thermal fluctuation effects beyond MF level on these predictions. As we shall see this is a surprisingly subtle problem and complete answers cannot be given for the renormalized phase boundary. Indeed, even at MF level problems arise since a simple effective Hamiltonian approach does not quite recover all the quantitative results of the Landau theory. In discussing these issues we also seek to explain why a previous effective Hamiltonian study \cite{rn} did not find any evidence for a corrugation-induced first-order wetting transition in the non-planar geometry.

In the last ten years or so one of the main thrusts in theoretical wetting research has been the study of adsorption in such non-planar geometries. The conformation of thin liquid films on rough surfaces has been described quite extensively \cite{andelman} while the influence of disordered (self-affine) substrates on three dimensional wetting transitions has also been investigated, using both replica and renormalization group methods \cite{kardar}. New critical behaviour is indicated when the roughness of the substrate exceeds the thermal roughness (as measured by the roughness exponent \cite{lf}) of the wetting layer. However, the transition remains second-order in character. In two dimensions, this no longer need be true and even richer behaviour occurs \cite{giug}. Work has also be carried out for systems with van der Waal's interactions for both the wedge \cite{dietrich2} and groove \cite{yeomans} geometries and quite general conditions for roughness-induced wetting have been found \cite{roland}.

To begin we recall our basic `microscopic' model of fluid adsorption at a corrugated wall (appropriate to systems with short-range forces) and discuss the results and interpretation of our earlier work. This is of some importance, as we shall show that the simplest available phenomenological approach does not quite agree with all our previous predictions.

\section{Perturbative analysis of a Landau theory}
Writing $z_W({\bf y})$, with ${\bf y}=(y_1,y_2)$, for the local height of the wall above the $z=0$ plane we consider a Landau-Ginzburg-Wilson (LGW) Hamiltonian of the order parameter (magnetization) $m({\bf r})$,
\eqan{HLGW}
H_{LGW}[m] &=& \int d{\bf y} \Biggl \{ \int_{z_W}^\infty dz \Biggl[ \half (\nabla_\perp m)^2 + \half \left( \frac{\partial m}{\partial z} \right)^2 \nonu \\
& & + \phi(m) \Biggr] + \left[ 1+\half (\nabla_\perp z_W)^2 \right] \phi_1(m_1) \Biggr \}
\naqe
where $m_1$ denotes the surface magnetization at vector position ${\bf y}$ along the $z=0$ plane, i.e.\ $m_1({\bf y}) = m({\bf y},z_W({\bf y}))$, while $\phi(m)$ and $\phi_1(m)= \half c m^2 - h_1 m$ are appropriate bulk and surface potentials. Here $c$ and $h_1$ are the surface enhancement and field, respectively. The final multiplicative term in (\ref{HLGW}) represents the local increase in area due to the corrugations of the wall expanded to second-order in $z_W$. Note, we have written $\nabla = \left( \n,\frac{\partial}{\partial z} \right)$ with $\n=\left( \frac{\partial}{\partial y_1},\frac{\partial}{\partial y_2} \right)$. We suppose that the bulk exhibits two-phase coexistence in zero bulk field ($h=0$) between phases with magnetization $m_\alpha$ ($>0$) and $m_\beta$ ($<0$) for $T<T_C$. Furthermore, we set $h_1<0$ and focus on wetting of the wall-$\alpha$ interface by the $\beta$ phase.

At mean-field level, corresponding to Landau or square gradient theory, the Hamiltonian needs to be minimized to obtain the equilibrium magnetization. This is a straightforward task for the planar system $z_W=0$. Denoting $\kappa = \sqrt{\phi''(m_\beta)}$, the inverse bulk correlation length of the $\beta$ phase, we recall that for $c$ greater (less) than $\kappa$ the wetting transition is second (first)-order \cite{telo}. For the non-planar system, however, even the MF calculation is non-trivial due to the loss of translational invariance along the wall. Previously \cite{us}, we have developed a perturbative approach to the problem in which the equilibrium free energy of the wall-$\alpha$ interface is written 
\eqan{prF}
{\cal F} &=& \phi(m_\alpha) V_\pi + \sigma_{w \alpha} A_\pi \nonu \\
& & + \frac{1}{2 (2 \pi)^{(d-1)}} \int d{\bf q} \; q^2 \Delta_\pi({\bf q}) |\hat{z}_W({\bf q})|^2 + \cdots
\naqe
Here the first two contributions represent the bulk and surface free energy (tension) of the appropriate planar system while the final term is the non-planar correction written perturbatively in terms of the Fourier amplitudes of the wall function $z_W({\bf y})$. The quantity to be determined $\Delta_\pi({\bf q})$ has the dimensions of a surface tension and by construction satisfies the long wavelength identity
\equn{shalom}
\Delta_\pi(0) = \sigma_{w\alpha}
\nuqe
required from infinitesimal rotational invariance. More generally, $\Delta_\pi({\bf q})$ can be related to the surface magnetization $m_{\pi1}$ and correlation functions of the planar system
\equn{prF2}
q^2 \Delta_\pi({\bf q}) = \phi_1(m_{\pi 1}) + m_{\pi 1}^{'2} \left( \frac{1}{\hat{G}(0,0;{\bf q})} - \frac{1}{\hat{G}(0,0;{\bf 0})} \right)
\nuqe 
Here $\hat{G}(z_1,z_2;{\bf q}) =\hat{G}(z_1,z_2;q)$ is the transverse Fourier transform of the connected two-point correlation function, $G({\bf r}_1,{\bf r}_2) = \langle m({\bf r}_1) m({\bf r}_2) \rangle - \langle m({\bf r}_1) \rangle \langle m({\bf r}_2) \rangle$, for positions distances $z_1$ and $z_2$ from the wall, i.e.\ it is surface correlations which determine the free energy correction. In this way one may derive an elegant relation for $\Delta_\pi({\bf q})$ ($= \Delta_\pi(q)$ from (\ref{prF2})) which complements the well known expression for the surface tension \cite{telo}
\equ
\sigma_{w\alpha} = \phi_1(m_{\pi 1}) + \int_{m_{\pi 1}}^{m_b} dm Q_0(m)
\uqe 
where $m_b= m_\pi(\infty)$ is the bulk magnetization and $Q_0(m) = \sqrt{2(\phi(m) - \phi(m_b))}$ is the usual function appearing in the Cahn construction for planar square gradient theories. The final expression for $\Delta_\pi(q)$ is
\equ
\Delta_\pi(q) = \phi_1(m_{\pi 1}) + \int_{m_{\pi 1}}^{m_b} dm Q(m;q)
\uqe
where $Q(m;q)$ is a modified Cahn function satisfying the differential equation
\equ
\frac{d}{dm} \left[ Q_0^3 \frac{d}{dm} \left( \frac{Q}{Q_0} \right) \right] = q^2 Q
\uqe
Clearly, when $q=0$ we have $Q(m;0) = Q_0$ and $\Delta_\pi$ reduces to $\sigma_{w\alpha}$ as quoted earlier.

Application of this perturbation theory to the case of adsorption at a corrugated sinusoidal wall, 
\equn{zw}
z_W({\bf y}) = \sqrt{2} A \sin (p y_1)
\nuqe
with $A$ the root mean square width of the corrugations and $2 \pi/p$ their period, yields results best expressed in terms of the contact angle $\theta_\pi = \theta_\pi(T,c,h_1)$ of the planar system. Thus, for the case of strongly first-order wetting transitions in the planar system occurring at temperature $T_\pi$, say (for which $\theta_\pi \propto (T_\pi-T)^\half$), it was found that the corrugated wall-$\alpha$ interface was completely wet by the $\beta$ phase at a {\it lower} temperature satisfying 
\equn{fo}
\theta_\pi = p A \mbox{\hspace*{2cm} for $c<\kappa$}
\nuqe
(valid for $p \ll \kappa$) with the transition remaining first-order. Indeed, the possibility of corrugation-induced second-order wetting transitions is ruled out completely from the expansion (\ref{prF}) with (\ref{prF2}).

On the other hand, for planar second-order wetting transitions (for which $\theta_\pi \propto (T_\pi-T)$) the corrugated geometry showed a first-order phase transition for sufficiently large $A>A^*$ at a reduced temperature satisfying
\equ
\theta_\pi \approx p \left[ A^2 - A^{*2} \right]^\half \mbox{\hspace*{2cm} for $c>\kappa$}
\uqe
For $A<A^*$ the transition remained second-order and occurred at the same planar wetting temperature $T_\pi$. Surprisingly, the threshold or tricritical amplitude $A^*$ is comparatively small and is explicitly determined as
\equn{itsa*}
\kappa A^* = \sqrt{\frac{c-\kappa}{c+\kappa}} \mbox{\hspace*{2cm} for $c>\kappa$}
\nuqe
which is less than a bulk correlation length. Thus, even minor deviations from the plane can lead to a corrugation-induced first-order wetting transition.

Before trying to rederive these results using an effective Hamiltonian approach it is worth mentioning a few points concerning their interpretation. Firstly, the first-order result (\ref{fo}) is precisely the expression obtained from a naive application of Wenzel's empirical law \cite{wenzel} to wetting transitions. Recall that Wenzel observed that the contact angle $\theta_\rho$ of a droplet on a rough surface (of area $A_\rho$) appeared to satisfy the relation 
\equn{wenzel}
\frac{\cos \theta_\rho}{\cos \theta_\pi} = \frac{A_\rho}{A_\pi}
\nuqe
For a corrugated wall $\frac{A_\rho}{A_\pi} = 1 + \half p^2 A^2$ (to quadratic order in $pA$) and setting $\theta_\rho = 0$ recovers (\ref{fo}) for small $\theta_\pi$. However, the macroscopic Wenzel law (\ref{wenzel}) is certainly not universally valid and is restricted to adsorption problems in which the transverse correlation length characterizing the capillary wave-like fluctuations of the $\alpha \beta$ interface is not much larger than the bulk correlation length. This condition is met for strongly first-order phase transitions since the thickness of the wetting $\beta$ layer does not exceed a few bulk correlation lengths before the transition to infinite adsorption occurs.

\begin{figure}[h]
\begin{center}
\scalebox{0.4}{\includegraphics{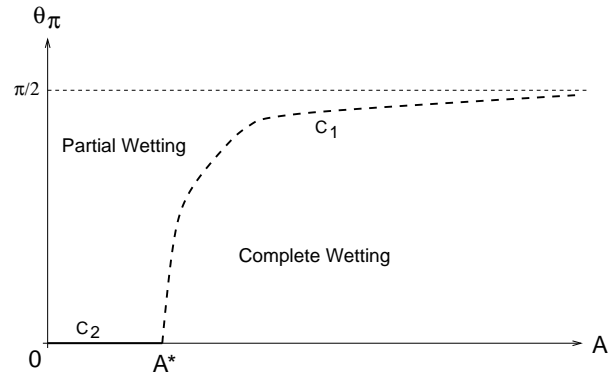}}
\end{center}
\caption{Schematic wetting phase diagram for fluid adsorption in a system with a corrugated wall and $c>\kappa$. The lines $C_1$ and $C_2$ are loci of first and second-order wetting phase transitions, respectively, which meet at the tricritical point corresponding to $A=A^*$. The vertical axis is a linear measure of the temperature scale $T_\pi -T$ for small $\theta_\pi$, where $T_\pi$ is the wetting temperature in the planar system.}
\label{rwpd}
\end{figure}

\begin{figure}[h]
\begin{center}
\scalebox{0.4}{\includegraphics{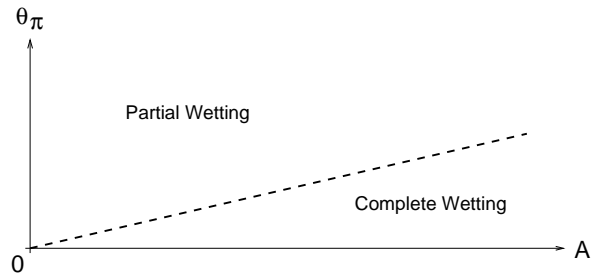}}
\end{center}
\caption{For a system with $c<\kappa$ only first-order transitions take place. The wetting temperature is reduced below $T_\pi$ by an amount proportional to the wall amplitude squared (at least for small $A$).}
\label{rwpd2}
\end{figure}

Secondly, for $c>\kappa$ (i.e.\ planar second-order wetting transitions) the amplitude $A^*$ vanishes smoothly as the planar tricritical point $c=\kappa$ is approached. We emphasize here that this is required in order that the global surface phase diagram shows a smooth cross-over to the Wenzel-like result (\ref{fo}) appropriate for the planar first-order regime. Given that we can discount the possibility of corrugation-induced second-order behaviour in the $c<\kappa$ sector, a non-vanishing value of $A^*$ as $c \goto \kappa^+$ would somewhat surprisingly result in a discontinuous surface phase diagram. Consequently, the prediction (\ref{itsa*}) imbues the surface phase diagram with a natural topology facilitating a smooth cross-over near planar tricriticality. Sections through this diagram are sketched in Fig.\ \ref{rwpd} and Fig.\ \ref{rwpd2}.

Finally, we note that deep in the planar second-order sector $c \gg \kappa$ the result for $A^*$ exhibits universal-like properties
\equn{uni}
\kappa A^* \approx 1 \mbox{\hspace*{2cm} for $c \gg \kappa$}
\nuqe
This prediction is appropriate for planar critical wetting transitions close to the bulk critical temperature and sufficiently far from the planar tricritical point. As we shall see, the right hand side of this equality is associated with the numerical value of a hyperscaling amplitude.

\section{Effective Hamiltonian Theory}
\subsection{Predictions of a simple model}
The simplest continuum interfacial Hamiltonian model of fluid adsorption at a non-planar wall is 
\equn{Hi}
H[\ell,z_W] \approx \int d{\bf y} \Biggl \{ \half \Sigma_{\ab} (\n \ell)^2+ W(\ell-z_W) \Biggr \}
\nuqe
where $\Sigma_{\alpha \beta}$ is the stiffness coefficient of the unbound $\alpha \beta$ interface, $W(\ell)$ is the binding potential for adsorption at a planar wall and $\ell({\bf y})$ is a suitable measure of the local thickness of the wetting film. This is certainly a plausible starting point for investigations and has been employed by numerous authors \cite{vol14} to understand interfacial fluctuation effects at rough and self-affine walls. Therefore, it is surprising to note that the model does not recover all the Landau MF results quoted earlier. Nevertheless, it does manage to capture the correct physics away from the planar tricritical point and provides a simple means of understanding the origin of the corrugation-induced first-order behaviour for $c \gg \kappa$. Before discussing the effect of wall corrugation on planar first- and second-order wetting transitions we make a few general remarks. Firstly, at MF level the problem reduces to finding the global minimum of $H[\ell,z_W]$. A wetting phase transition occurs when the global minimum approaches zero from below so that the free energy is identical to that of an unbound interface $H[\infty,z_W]=0$ which always represents a minimum. This comparison of free energies appears not to have been adopted by Rejmer and Napi\'{o}rkowski \cite{rn} (RN) in their effective Hamiltonian study and as a consequence they concentrate solely on second-order behaviour (note these authors actually consider a more complicated effective Hamiltonian model systematically derived from the LGW functional). Secondly, following standard theory, the MF equilibrium values of the planar film thickness $\ell_\pi$, correlation length $\xi_\parallel$ and singular free energy $f^{(\pi)}_{\rm sing}$ are identified through
\equn{lots}
\begin{array}{ccccc}
W'(\ell_\pi) = 0 &;& \xi_\parallel^2 = \frac{\Sigma_{\alpha \beta}}{W''(\ell_\pi)} &;& f^{(\pi)}_{\rm sing} = W(\ell_\pi)
\end{array}
\nuqe 
The latter quantity is conveniently related to the planar contact angle by Young's equation \cite{dietrich}
\eqa
f^{(\pi)}_{\rm sing} &=& \sigma_{\alpha \beta} (\cos \theta_\pi -1) \nonu \\
&\approx & \half \sigma_{\alpha \beta} \theta_\pi^2
\aqe
where $\sigma_{\alpha \beta}$ is the surface tension of the $\alpha \beta$ interface. Denoting $\ell_\rho({\bf y})$ the collective coordinate field that minimizes $H[\ell,z_W]$ we have
\equn{EL}
\Sigma_{\ab} \n^2 \delta \ell_\rho = W'( \delta \ell_\rho - z_W)
\nuqe
which has to be solved perturbatively. Assuming that $|z_W({\bf y})|$ and $\delta \ell_\rho = \ell_\rho - \ell_\pi$ are small, we find to first order
\equ
\Sigma_{\ab} \n^2 \delta \ell_\rho = W''(\ell_\pi) \Bigl( \delta \ell_\rho - z_W \Bigr)
\uqe
which is trivially solved on introducing the Fourier transforms
\eqa
\delta \hat{\ell}_\rho ({\bf q}) &=& \int d{\bf y} {\rm e}^{i {\bf q}.{\bf y}} \delta \ell_\rho({\bf y}) \\
\hat{z}_W ({\bf q}) &=& \int d{\bf y} {\rm e}^{i {\bf q}.{\bf y}} z_W({\bf y})
\aqe
We find
\equn{ELsol}
\delta \hat{\ell}_\rho({\bf q}) = \frac{\hat{z_W}({\bf q})}{1+q^2 \xi_\parallel^2}
\nuqe
and see that the `healing length', in the terminology of \cite{andelman}, is equal to the transverse correlation length of the planar system. Short wavelength surface undulations, such that $q \xi_\parallel \gg 1$, are completely damped by the interfacial stiffness. The interface is flat and  the effects of surface roughness are negligible. We also point out that it is possible to go beyond the simple form of the effective Hamiltonian given in (\ref{Hi}) and to include a non-local interaction between the wall and the $\alpha \beta$ interface. Equation (\ref{ELsol}) is then modified by the appearance of a kernel function which is simply unity for our case (see \cite{andelman} for more details).

Thus to quadratic order in $\hat{z}_W({\bf q})$ (equivalent to the free energy expansion (\ref{prF}) of the Landau theory), the singular contribution to the surface free energy for a bound (non-wet) phase at a corrugated wall is 
\eqa
F_{\rm non-wet} &=& H[\ell_\rho,z_W] \nonu \\
&=& A_\pi W(\ell_\pi) \nonu \\
& & + \frac{1}{2 (2 \pi)^2} \int d{\bf q} \frac{\Sigma_{\alpha \beta} q^2}{1 + q^2 \xi_\parallel^2} |\hat{z}_W({\bf q})|^2
\aqe
which must be compared with
\eqa
F_{\rm wet} &=& H[\infty,z_W] \nonu \\
&=& 0
\aqe
If we specialize to the case of a corrugated wall (\ref{zw}), the integral can be evaluated and the wetting phase boundary is then given by the solution of 
\equn{solution}
|W(\ell_\pi)| = \frac{\Sigma_{\alpha \beta} A^2 p^2}{2(1+p^2 \xi_\parallel^2)}
\nuqe

\subsubsection{First-order planar wetting transitions}
At a first-order transition the singular free energy $f_{\rm sing}$ vanishes linearly as $T \goto T_\pi^-$ and the correlation length $\xi_\parallel$ remains finite at the transition. Thus, provided the wavelength $p^{-1}$ of the corrugations is much larger than the transverse correlation length of the (planar) thin film, we find a shifted first-order wetting transition occurring at a lower temperature
\equn{wlike}
\theta_\pi \approx \sqrt{\frac{\Sigma_{\alpha \beta}}{\sigma_{\ab}}} A p \mbox{\hspace*{2cm} for $p \xi_\parallel \ll 1$}
\nuqe
identical to (\ref{fo}) for isotropic fluid interfaces (and recall $\theta_\pi \sim (T_\pi-T)^\half$). Close to the planar tricritical point $c=\kappa$, where $\xi_\parallel$ is large for the thin film phase, this prediction is no longer accurate and (\ref{solution}) suggests that the expression for the shifted phase boundary shows cross-over to $\theta_\pi \propto \frac{A}{\xi_\parallel}$ for fixed $p$. However, we will not dwell on this since, as we shall show below, the simple interfacial model is somewhat unreliable close to planar tricriticality ($c \approx \kappa$).

\subsubsection{Second-order planar wetting transitions}
The simple Wenzel-like result (\ref{wlike}) is clearly inapplicable for describing the effect of corrugation on a planar second-order wetting transition due to the divergence of the correlation length $\xi_\parallel$ as $T \goto T_\pi^-$. Instead, directly from (\ref{solution}), we have
\equn{gotya}
\frac{\sigma_{\ab}}{\Sigma_{\ab}} \theta_\pi^2 = A^2 p^2 + 2 \frac{ f^{(\pi)}_{\rm sing} \xi_\parallel^2}{\Sigma_{\ab}} p^2
\nuqe
for the shifted phase boundary corresponding to a corrugation-induced first-order transition. In fact, the absence of real solutions (for $\theta_\pi$) in (\ref{gotya}) indicates that the transition remains second-order. This expression further simplifies because the hyperscaling amplitude ratio
\equn{Rdef}
{\cal R} = \; \stackrel{\rm lim}{\scs T \goto T_\pi} \left\{ -\frac{f_{\rm sing}^{(\pi)} \xi_\parallel^2}{(\Sigma_{\ab} / \kappa^2)} \right\}
\nuqe
is a pure number in MF theory (see below). Accepting this for the moment, we conclude that the interfacial model predicts a shifted wetting phase boundary of the form
\equn{pbound}
\theta_\pi = \left\{ \begin{array}{lr} p \left[ \frac{\Sigma_{\ab}}{\sigma_{\ab}} (A^2 - A^{*2}) \right]^\half & \mbox{for $A>A^*$} \nonu \\
0 & \mbox{for $A<A^*$} \end{array} \right.
\nuqe
and recall $\theta_\pi \sim (T_\pi-T)$. Moreover, the tricritical amplitude $A^*$ is simply determined by the value of the amplitude ratio ${\cal R}$,
\equn{Rres}
\kappa A^* = \sqrt{2 {\cal R}}
\nuqe

At this stage the results of the interfacial model appear very similar to those of the Landau theory calculation for which we can identify $\sigma_{\ab} = \Sigma_{\ab}$. However, the prediction for $A^*$ does not quite agree as seen by calculating the value of ${\cal R}$. If we adopt the standard form for the binding potential in zero field ($h=0$)
\equn{w}
W(\ell) = \left\{ \begin{array}{lcl} -a \e^{-\kappa \ell} + b \e^{-\kappa \ell} & & \mbox{for $\ell>0$} \nonu \\
\infty & & \mbox{for $\ell<0$} \end{array} \right.
\nuqe
with $a \sim T_\pi - T$ and $b>0$, a simple calculation yields ${\cal R} = \half$ so that
\equn{itsalsoa*}
\kappa A^* = 1
\nuqe
instead of the result (\ref{itsa*}). Unlike the direct analysis of the Landau theory functional, the simple interfacial model predicts that the amplitude $A^*$ is independent of the surface enhancement $c$ and does not vanish as the planar tricritical point ($c=\kappa$) is approached. Nevertheless, this approach does agree with the results of our earlier theory for strongly first- and second-order (planar) wetting transitions corresponding to $c \ll \kappa$ and $c \gg \kappa$, respectively. In particular, the `universal' result (\ref{uni}) can be traced to the numerical value of the appropriate ratio ${\cal R}$ of hyperscaling amplitudes.

Before discussing the possible reasons for the discrepancy between the Landau and interfacial model calculations close to planar tricriticality we focus on the strongly second-order sector $c \gg \kappa$ and enquire how results (\ref{pbound}) and (\ref{Rres}) are modified by thermal fluctuations in three dimensions.

\subsection{Fluctuation effects away from planar tricriticality}
Wall corrugation has only a minor influence on wetting transitions below the upper critical dimension where entropic fluctuation effects dominate the unbinding mechanism \cite{us2}. However, at the upper critical dimension $d=3$ (restricting our attention to systems with short-range forces) we can anticipate that many of the qualitative features seen in the MF calculation retain relevance even after thermal fluctuations are allowed for. If we simply assume that the phenomenological model (\ref{Hi}) is a reasonable description of fluctuation effects ({\it away} from the planar tricritical point) it is straightforward to develop a linear functional renormalization group (RG) analysis along the lines formulated by Fisher and Huse (FH) \cite{fh} for the planar problem $z_W=0$. In fact, assuming that the renormalized Hamiltonian $H^{(t)}[\ell,z_W]$ is of the same functional form as (\ref{Hi}), then the RG transformations are unchanged and
\equn{Hi2}
H^{(t)}[\ell,z_W] = \int d{\bf y} \left\{ \half \Sigma_{\ab} (\n \ell)^2 + \e^{-2t} W^{(t)}(\ell-z_W) \right\}
\nuqe
in $d=3$, where $b=\e^t$ is the usual spatial rescaling factor and 
\equ
W^{(t)}(\ell-z_W) = \frac{{\rm e}^{2t}}{\sqrt{4\pi \omega t}} \int_{-\infty}^\infty d\ell' W(\ell') {\rm e}^{-\frac{(\ell-z_W-\ell')^2}{4\omega t}}
\uqe
Here $\omega$ is the wetting parameter
\equ
\omega = \frac{k_B T \kappa^2}{4 \pi \Sigma_{\ab}}
\uqe
with $k_B$ the Boltzmann constant. Following FH we adopt a matching procedure and choose $t=t^*$ such that the renormalized curvature $W^{(t)''}(\ell)$ is about unity (see appendix) at its minimum. The structure of the perturbative analysis of (\ref{Hi2}) is then identical to that described earlier for MF theory (equations (\ref{EL})--(\ref{solution})). Consequently, the expression for the shifted phase boundary and threshold corrugation amplitude are identical to those quoted earlier (see (\ref{pbound}) and (\ref{Rres}), respectively) although the temperature dependence of $\theta_\pi$ and the numerical value of ${\cal R}$ are now dependent on $\omega$.

Thus, the planar contact angle vanishes like
\equ
\theta_\pi \sim (T_\pi-T)^{\nu_\parallel}
\uqe
where \cite{fh}
\equ
\nu_\parallel = \left\{ \begin{array}{lcl} \frac{1}{1-\omega} & & \mbox{for $0<\omega <\half$} \nonu \\
\frac{1}{(\sqrt{2}-\sqrt{\omega})^2} & & \mbox{for $ \half <\omega <2$}
\nonu \\
\infty & & \mbox{for $\omega > 2$}
\end{array} \right.
\uqe
is the standard correlation length critical exponent describing the divergence of $\xi_\parallel$ in the planar system as $T \goto T_\pi^-$. Furthermore, one can determine the value of the amplitude ratio ${\cal R}^{(t)}$ (see appendix) so that the renormalized expression for the threshold amplitude $A^*$ satisfies
\equ
\kappa A^* \approx \left\{ \begin{array}{lcl} 1 & & \mbox{for $0<\omega<\half$} \nonu \\
(2 \omega)^\frac{1}{4} & & \mbox{for $\omega > \half$} \\
\end{array} \right. 
\uqe
which we emphasize must be regarded as a `high' temperature prediction valid away from the planar tricritical point. Thus, the effect of thermal fluctuations is to increase the threshold amplitude $A^*$ so that it is more difficult to produce a corrugation-induced first-order transition. This is supported by the exact results found in $d=2$ where the transition is {\it always} second order independent of the size of $A$ \cite{us2}. Nevertheless, the numerical value of $A^*$ remains relatively small for physically relevant estimates of $\omega$. For example, close to the bulk critical temperature $\omega$ approaches a universal value $\omega_C \approx 0.77$ \cite{fw} which implies that $\kappa A^* \approx 1.1$ if $T_\pi$ is close to $T_C$. This is hardly different from the MF prediction and could be checked in Ising model simulation studies.

Finally, in this subsection we note that in deriving these results we have taken as our starting point the simplest phenomenological model of fluctuation effects at wetting transitions and have ignored the possibility of including a position dependent stiffness coefficient \cite{jin} and also coupling to order parameter fluctuations near the wall \cite{2f,2f2,bigbefore,big}. While both these amendments to the standard model have some important consequences in $d=3$ they do not effect the location of the planar wetting phase boundary to any great extent. Thus, while the current theoretical expectation \cite{chris} is that the planar wetting transition (occurring in the Ising model say) is extremely weakly first-order beyond MF level (even for $c>\kappa$) there is no direct evidence for this in Ising model simulation studies which certainly appear to show second-order behaviour \cite{parryrev}. Consequently, we feel justified in ignoring the possibility of a fluctuation-induced first-order transition since it is unlikely to interfere with the mechanism for corrugation-induced first-order wetting which is our central concern here.

\subsection{Remarks on improved effective Hamiltonians}
To complete our article we return to the discrepancy between the perturbative Landau theory and simple interfacial model predictions for the tricritical amplitude $A^*$ (equations (\ref{itsa*}) and (\ref{itsalsoa*}), respectively). Both results agree for $c \gg \kappa$ and are consistent with independent numerical minimization of the Landau free energy functional in this limit \cite{tobe}. However, they are qualitatively different for $c \approx \kappa$ since only within the perturbative Landau theory calculation does $A^*$ vanish, as it must, as planar tricriticality is approached. For, as mentioned earlier, a non-vanishing $A^*$ would imply a discontinuous cross-over to the Wenzel-like behaviour expected in the first-order sector $c<\kappa$. 

In principle, it is of course possible to recover all the Landau theory using an effective Hamiltonian. If we denote $\ell({\bf y})$ as the surface of fixed magnetization $m^X=0$ then a constrained minimization of the LGW functional subject to the crossing constraint $m({\bf y},\ell({\bf y}))=0$ (see \cite{jin}) defines a Hamiltonian 
\equn{effH}
H[\ell,z_W] = \min H_{LGW}[m]
\nuqe
which identically must recover the MF free energy (\ref{prF}) on further minimization with respect to the collective coordinate $\ell({\bf y})$. However, the constrained minimization involved in (\ref{effH}) is, of course, at least as difficult as the original MF theory of (\ref{HLGW}) unless various approximations are used \cite{jin}. It is these very approximations which lead to the discrepancy in $A^*$ described above.

In fact, one should not be surprised at the limited domain of validity of  (\ref{Hi}). As is now appreciated even for a planar substrate, the simple interfacial model does not provide an accurate description of magnetization correlations near the wall \cite{bigbefore,big} but it is these very correlations that are explicitly incorporated into the free energy correction kernel $\Delta_\pi(q)$. To derive the correct Landau expression for the wall correlation function $\hat{G}(0,0;q)$ a two-field Hamiltonian $H[X,\ell]$ \cite{big} is required, where $X({\bf y})$ is the collective coordinate most suitable for modelling magnetization fluctuations near the surface. Note, that this need not be interfacial-like and indeed is not so at the critical wetting transition \cite{2f2,big}.

This suggests that the effective Hamiltonian for wetting at a non-planar wall should resemble the two-field models of wetting at a planar surface, provided $X({\bf y})$ is chosen to be an interfacial-like variable, describing translations of a contour of fixed magnetization close to the value at the wall. This is indeed the case and the form for $H[\ell,z_W]$ derived by RN is almost identical to the two-field Hamiltonian of Boulter and Parry \cite{2f}. Using the notation of the latter the generic form for $H[\ell,z_W]$ in the long wavelength limit is
\eqan{twof}
H[\ell,z_W] &=& \int d{\bf y} \biggl\{ \half \Sigma_{11}(\ell-z_W) (\n z_W)^2 \nonu \\
& & + \Delta \Sigma_{12}(\ell-z_W) \n \ell. \n z_W \nonu \\
& & + \half \Sigma_{22}(\ell-z_W) (\n \ell)^2 + W(\ell-z_W) \biggr\}
\naqe
where the $\Sigma_{\mu \nu}$ constitute the elements of a stiffness-matrix describing the position dependent corrections to the separate surface tensions
\eqa
\Sigma_{11} = \sigma_{w\alpha} + \Delta \Sigma_{11}(\ell-z_W) \\
\Sigma_{22} = \sigma_{\ab} + \Delta \Sigma_{22}(\ell-z_W)
\aqe
All the $\Delta \Sigma_{\mu \nu}(\ell)$ vanish as $\ell \goto \infty$ and if this position dependence is ignored (\ref{twof}) becomes simply (\ref{Hi}). From the explicit expressions for the various functions, RN derive the following asymptotic expansions (in zero field $h=0$)
\eqan{a}
\Delta \Sigma_{11}(\ell) &=& u_{10} X + (u_{20} + u_{21} \kappa \ell) X^2 + \cdots \\
\Delta \Sigma_{12}(\ell) &=& w_{10} X + (w_{20} + w_{21} \kappa \ell) X^2 + \cdots \label{b} \\
\Delta \Sigma_{22}(\ell) &=& (\eta_{10}+\eta_{11} \kappa \ell) X + (\eta_{20} + \eta_{21} \kappa \ell) X^2 + \cdots \label{c} \\
W(\ell) &=& v_{10} X + v_{20} X^2 + \cdots \label{d}
\naqe
where $X = \e^{-\kappa \ell}$ and the ellipses denote terms of cubic and higher order in $X$. 

Here we point out, for the first time, that the four functions $\Delta \Sigma_{\nu \mu}(\ell)$ and $W(\ell)$ are not independent but necessarily satisfy a functional stiffness-matrix binding potential relation
\equn{SMFE}
\Delta \Sigma_{11}(\ell) + 2 \Delta \Sigma_{12}(\ell) + \Delta \Sigma_{22}(\ell)  = W(\ell) - \ell W'(\ell)
\nuqe
which is the analogue of the identity (\ref{shalom}) in the Landau perturbation theory. This relation is valid for arbitrary choices of the potentials $\phi(m)$ and $\phi_1(m)$, and if $H[\ell,z_W]$ is defined via a partial trace \cite{jin} rather than just a MF saddle point identification (as in (\ref{effH}) and \cite{rn}). The identity follows from simply requiring the Hamiltonian (\ref{twof}) to be invariant with respect to infinitesimal rotations of the plane of the wall. For the equilibrium planar MF position $\ell_\pi$ satisfying $W'(\ell_\pi)=0$, the identity reduces to the (bare) stiffness-matrix free energy relation pertinent to the two-field theory of wetting \cite{bigbefore}. Equation (\ref{SMFE}) is a more general requirement and significantly constrains the behaviour of the coefficients in the asymptotic expansions (\ref{a}--\ref{d}). Specifically, we obtain the following 
\eqan{simple}
v_{10} &=& 2 \eta_{11} \\
v_{10} &=& w_{10} + u_{10} + 2 \eta_{10} \\
v_{20} &=& w_{20} + u_{20} + 2 \eta_{20} \\
2 v_{20} &=& w_{21} + u_{21} + 2 \eta_{21}
\naqe
All these relations are obeyed by the coefficients explicitly calculated by RN using the double parabola approximation \cite{lipowsky}, but are equally valid for arbitrary $\phi(m)$.

Recall that a two-field approach seems sensible due to the appearance of the planar wall correlation function in (\ref{prF2}). As pointed out in \cite{bigbefore}, this correlation function can be recovered (in MF theory) by using a $H[\ell_1,\ell]$ Hamiltonian, with two interfacial fields modelling fluctuations at the wall and at the $\ab$ interface, respectively. However, such a tack is best suited for the complete wetting regime and at critical wetting $\ell_1$ should no longer have {\it any} interfacial-like component \cite{2f2,big} and as such a $H[X,\ell]$ Hamiltonian is the optimal choice. Unfortunately, the technique of `freezing' the lower field into the configuration of a rough wall (as used in (\ref{twof})) in order to describe wetting in a non-planar system is no longer valid. The field $X({\bf y})$ is not interfacial-like and consequently does not have the dimensions of length. Progress can be made by introducing {\it two} fields at the wall, one interfacial-like so that it can take up the wall configuration and the other non-interfacial-like to allow the Hamiltonian to recover critical wetting wall correlation functions in the limit of $z_W$ going to zero. However, pushing this three-field Hamiltonian beyond MF theory seems, at the moment, to be prohibitively difficult.

\section{Discussion}
In this paper we have rederived recent predictions for corrugation-induced first-order wetting transitions using an effective Hamiltonian (with a harmonic approximation to the binding potential). We have shown that the phase diagram (Fig.\ \ref{rwpd}) is valid in three dimensions and that the critical amplitude $A^*$, for which walls corrugated with $A>A^*$ discontinuously wet, depends on the wetting parameter $\omega$. The value of $A^*$ is related to a ratio of hyperscaling amplitudes --- a prediction which is open to investigation by Ising model simulation studies. Consistent with exact results in $d=2$ \cite{us2}, it can be seen that fluctuations extend the size of the second-order regime beyond the MF predictions. This extension is not very significant for the three dimensional Ising model at temperatures close to $T_C$.

Importantly, our results are not quantitatively reliable near planar tricriticality and we are only confident of the predicted values of $A^*$ for wetting transitions away from the region $c \approx \kappa$.

We discuss the form of an improved effective Hamiltonian as considered in \cite{rn}. We point out (as do RN) the similarity with the two-field theory of coupling effects at planar wetting transitions and indicate that an accurate description of surface correlations is needed for a global prediction of $A^*$. Our analysis is consistent with that of RN who did not compare the bound state free energy with that at $\ell = \infty$ and so did not explore the first-order regime. Nevertheless, they did show that any continuous divergence of the interface thickness necessarily occurs at the same wetting temperature as in the planar geometry. This is entirely in keeping with our earlier remark that for $A<A^*$ the transition remains second-order. Using a new stiffness-matrix binding potential relation we are also able to prove that the coefficients appearing in the Hamiltonian, via the asymptotic expansions of $\Delta \Sigma_{\mu \nu}$ and $W(\ell)$, are not independent but obey simple linear relations.

To conclude, we believe that corrugation-induced first-order wetting is present in three dimensional systems with short-range forces but that more work is required to elucidate the behaviour of the threshold amplitude $A^*$ near the planar tricritical point.

\appendix
\section*{Calculation of the renormalized amplitude ${\cal R}$}
As there is no non-trivial fixed point of the renormalization group representing the wetting transition, FH adopt (as mentioned previously) a matching procedure to determine the value of $t$ up to which renormalization takes place. This value $t=t^*$ is chosen to be that at which the transverse correlation length of the renormalized binding potential is of the same order as the non-critical bulk correlation length $\kappa^{-1}$. One requires
\eqan{RGc}
\left. \frac{\partial W^{(t^*)}(\ell)}{\partial \ell} \right|_{\breve{\ell}} & = & 0 \label{RGc1} \\
\left. \frac{\partial^2 W^{(t^*)}(\ell)}{\partial \ell^2} \right|_{\breve{\ell}} & = & \Sigma \kappa^2
\naqe
which defines the equilibrium wetting layer thickness $\breve{\ell}$ and allows a mean-field type analysis to be appropriate. By setting $\kappa=1$, equations (\ref{lots}), (\ref{Rdef}) and (\ref{Hi2}) imply that
\equn{ratio3}
{\cal R}^{(t)} = - \frac{W^{(t)}(\breve{\ell})}{{W^{(t)}}''(\breve{\ell})}
\nuqe
and so, similar to \cite{fh}, we have to consider three separate regimes for the three different possible forms of the binding potential.
\subsection*{Regime I $\omega < \half$}
For this case FH find that
\equ 
{\rm e}^{-2t} W^{(t)}(\ell) = -a \e^{\omega t - \ell} + b \e^{4 \omega t - 2 \ell}
\uqe
and so 
\equ
\e^{-\breve{\ell}} = \frac{a}{2b} \e^{-3\omega t}
\uqe
From (\ref{ratio3}), calculating ${\cal R}^{(t)}$ is straight-forward
\equ
{\cal R}^{(t)} = \frac{a^2/4b}{a^2/2b} = \half
\uqe
implying, via (\ref{Rres}), that $A^*$ not shifted from its mean-field value.

\subsection*{Regime II $\half < \omega < 2$}
The algebra is now a little more involved. The renormalized potential is approximately
\equ
\e^{-2t} W^{(t)}(\ell) = -a \e^{\omega t - \ell} + \frac{K}{\sqrt{t}} \e^{-\frac{\ell^2}{4 \omega t}}
\uqe
with $K$ a constant \cite{fh}. Imposing (\ref{RGc1}) we find
\equn{diff1}
a \e^{\omega t - \breve{\ell}} = \frac{\breve{\ell} K}{2 \omega t^\frac{3}{2}} \e^{-\frac{\breve{\ell}^2}{4 \omega t}}
\nuqe
while
\equ
\e^{-2t} W^{(t)''}(\breve{\ell}) \approx -a \e^{\omega t - \ell} + \frac{K \ell^2}{4 \omega^2 t^{\frac{5}{2}}} \e^{-\frac{\breve{\ell}^2}{4 \omega t}}
\uqe
Consequently,
\eqa
{\cal R}^{(t)} &=& - \frac{-a \e^{\omega t - \breve{\ell}}+\frac{K}{\sqrt{t}} \e^{-\frac{\breve{\ell}^2}{4 \omega t}}}{ -a \e^{\omega t - \ell} + \frac{K \ell^2}{4 \omega^2 t^{\frac{5}{2}}} \e^{-\frac{\breve{\ell}^2}{4 \omega t}}} \nonu \\
&=& \frac{2 \omega t}{\breve{\ell}}
\aqe
making use of (\ref{diff1}). However, FH show that
\equ
\breve{\ell} = \sqrt{8\omega} \left( t - \frac{1}{8} \ln t \right)
\uqe
and so
\eqan{reg2}
{\cal R}^{(t)} &=& \sqrt{\frac{\omega}{2}} \frac{1}{1-\frac{\ln t}{8t}} \nonu \\
&\approx& \sqrt{\frac{\omega}{2}} \label{reg22}
\aqe
for large $t$, i.e.\ $T$ close to $T_\pi$.

\subsection*{Regime III $\omega>2$}
For this case we adopt the notation of \cite{fh} and write $\ell=\mu t$, the binding potential has the form
\equ
W^{(t)}(\ell) = \frac{\e^{2t-\frac{\ell^2}{4\omega t}}}{\sqrt{4\pi \omega t}} K_t(\mu)
\uqe
where
\equn{K}
K_t(\mu) = \frac{-a}{1-\frac{\mu}{2\omega}} + \frac{b}{2-\frac{\mu}{2\omega}} + \frac{2\omega c}{\mu} + O(1/t)
\nuqe
The constant $c$ is introduced as the linear renormalization group cannot handle correctly a completely hard wall, see (\ref{w}), and so a `soft' approximation is used, with $W(\ell)=c$ for $\ell<0$. Note that this problem does not arise in certain non-linear formulations \cite{lf}.

The matching procedure leads to \cite{fh}
\eqan{reg3}
\breve{\ell} & \approx & \sqrt{8 \omega} t \label{reg3a} \\
K_\infty(a_C,\mu=\sqrt{8 \omega}) &=& 0
\naqe
where the latter defines the renormalized wetting temperature. To leading order
\equ
{W^{(t)}}''(\breve{\ell}) = \frac{\e^{2t-\frac{\breve{\ell}^2}{4\omega t}}}{\sqrt{4 \pi \omega t}} \frac{1}{t} \left(-\frac{\mu}{2\omega} \right) \frac{\partial K_t}{\partial \mu}
\uqe
implying that
\equ
{\cal R}^{(t)} = \frac{2 \omega t}{\mu} \frac{K_t}{K'_t}
\uqe
Writing $\tau \sim a-a_C \sim \frac{T_\pi-T}{T_\pi}$, FH find
\equ
\breve{\mu} \approx \sqrt{8\omega}+\tau + O(1/t)
\uqe
Equation (\ref{K}) can then be used to show that for $\mu \approx \breve{\mu}$
\eqa
K_t &=& O(\tau) + O(1/t) \nonu \\
K_t' &=& {\rm constant} + O(\tau) + O(1/t)
\aqe
Hence, if ${\cal R}^{(t)}$ is written as
\equ
{\cal R}^{(t)2} = \frac{2 \omega t}{\breve{\ell}} \left[ \frac{t K_t}{K_t'} \right]
\uqe
then the term in square brackets must be almost constant, let it be $C$, say, for $T$ near $T_\pi$, that is $\tau \ll 1$ and $t \gg 1$. Using (\ref{reg3a}), 
\equ
{\cal R}^{(t)} \approx C \sqrt{\frac{\omega}{2}}
\uqe 
with $C$ an unknown constant. However, by continuity of ${\cal R}^{(t)}$ at $\omega=2$ and comparing with (\ref{reg22}) we can see that $C=1$.

\end{document}